\documentclass[conference]{IEEEtran}
\IEEEoverridecommandlockouts
\usepackage{cite}
\usepackage{amsmath,amssymb,amsfonts}
\usepackage{algorithmic}
\usepackage{graphicx}
\usepackage{textcomp}
\usepackage{xcolor}
\usepackage{multirow}
\usepackage{array}
\usepackage{tikz}
\def\BibTeX{{\rm B\kern-.05em{\sc i\kern-.025em b}\kern-.08em
    T\kern-.1667em\lower.7ex\hbox{E}\kern-.125emX}}
\begin{document}

\title{Motor Imagery Classification Emphasizing Corresponding Frequency Domain Method based on Deep Learning Framework\\
\thanks{20xx IEEE. Personal use of this material is permitted. Permission
from IEEE must be obtained for all other uses, in any current or future media, including reprinting/republishing this material for advertising or promotional purposes, creating new collective works, for resale or redistribution to servers or lists, or reuse of any copyrighted component of this work in other works.}
\thanks{This work was partly supported by Institute of Information \& Communications Technology Planning \& Evaluation (IITP) grant funded by the Korea government (MSIT) (No. 2017-0-00432, Development of Non-Invasive Integrated BCI SW Platform to Control Home Appliances and External Devices by User’s Thought via AR/VR Interface; No. 2017-0-00451, Development of BCI based Brain and Cognitive Computing Technology for Recognizing User’s Intentions using Deep Learning; No. 2019-0-00079, Artificial Intelligence Graduate School Program (Korea University)).}

}

\author{\IEEEauthorblockN{Byoung-Hee Kwon}
\IEEEauthorblockA{\textit{Dept. Brain and Cognitive Engineering} \\
\textit{Korea University}\\
Seoul, Republic of Korea \\
bh\_kwon@korea.ac.kr}
\and
\IEEEauthorblockN{Byeong-Hoo Lee}
\IEEEauthorblockA{\textit{Dept. Brain and Cognitive Engineering} \\
\textit{Korea University}\\
Seoul, Republic of Korea \\
bh\_lee@korea.ac.kr}
\and
\IEEEauthorblockN{Ji-Hoon Jeong}
\IEEEauthorblockA{\textit{Dept. Brain and Cognitive Engineering} \\
\textit{Korea University}\\
Seoul, Republic of Korea \\
jh\_jeong@korea.ac.kr}}

\maketitle

\begin{abstract}
The electroencephalogram, a type of non-invasive-based brain signal that has a user intention-related feature provides an efficient bidirectional pathway between user and computer. In this work, we proposed a deep learning framework based on corresponding frequency empahsize method to decode the motor imagery (MI) data from 2020 International BCI competition dataset. The MI dataset consists of 3-class, namely `Cylindrical', `Spherical', and `Lumbrical'. We utilized power spectral density as an emphasize method and a convolutional neural network to classify the modified MI data. The results showed that MI-related frequency range was activated during MI task, and provide neurophysiological evidence to design the proposed method. When using the proposed method, the average classification performance in intra-session condition was 69.68\% and the average classification performance in inter-session condition was 52.76\%. Our results provided the possibility of developing a BCI-based device control system for practical applications.
\end{abstract}


\section{Introduction}
Brain-computer interface (BCI) system provides bidirectional pathway between user and computers using brain signals \cite{vaughan2003brain}. Electroencephalogram (EEG) is a type of non-invasive-based signal that has advantage of higher time resolution than other signals such as functional magnetic resonance imaging and near-infrared spectroscopy \cite{chen2016high, lee2017network}. Therefore, EEG-based BCI devices enables direct communication with low latency between users and computers using brain signals that contain user intention.

Many related studies \cite{jeong2019trajectory, he2018brain, lee2020continuous, park2016movement, suk2014predicting} used the user's movement intentions and the various BCI paradigms such as steady-state visual evoked potential (SSVEP) \cite{kwak2017convolutional, chen2018control, won2015effect}, event-related potentials (ERPs) \cite{yeom2014efficient, lee2018high}, and motor imagery (MI) \cite{kim2014decoding, ofner2014using, kam2013non, won2017motion}. Among these paradigms, SSVEP and ERPs are the exogenous paradigms that require additional external devices. The additional devices interfere with concentration when users control BCI-based devices. To overcome these inconvenience, many previous studies focused MI paradigm, one of the endogenous paradigms that does not require external stimulus to control BCI-based devices such as robotic arm\cite{meng2016noninvasive, shiman2017classification, spataro2017reaching} and wheel chair \cite{kim2016commanding}. 

MI has significant brain signal features similar to real movement without the user's actual movement. The user can constitute a mentally rehearsed task intuitively by imagining the movement of muscles. For different movement imaginations that reflect users' intentions, brain signals induce different spatial and temporal frequency information. During MI task, event-related desynchronization/synchronization (ERD/ERS) features are revealed in beta band [14-30] Hz and mu band [8-14] Hz respectively \cite{jeong2020brain, korik2018decoding}, which are referred to as a sensory-motor rhythm in the primary sensorimotor area. Based on this neurophysiological basis, we considered that MI is suitable for controlling BCI-based devices that reflect user intentions.

The non-stationary properties are challenging issues in decoding the human brain signals \cite{lee2015subject}. Since the brain signals such as EEG are generated from different sources, non-stationary properties occur. These non-stationary properties cause variations and shifts in the EEG signals and these phenomena lead to a decrease of decoding performance in session-to-session and inter-subject transfers. To overcome the non-stationary properties related issue, many researchers have proposed various methods \cite{zhang2019survey, raza2019covariate}.

\begin{figure*}[t]
\centering
\includegraphics[width=\textwidth]{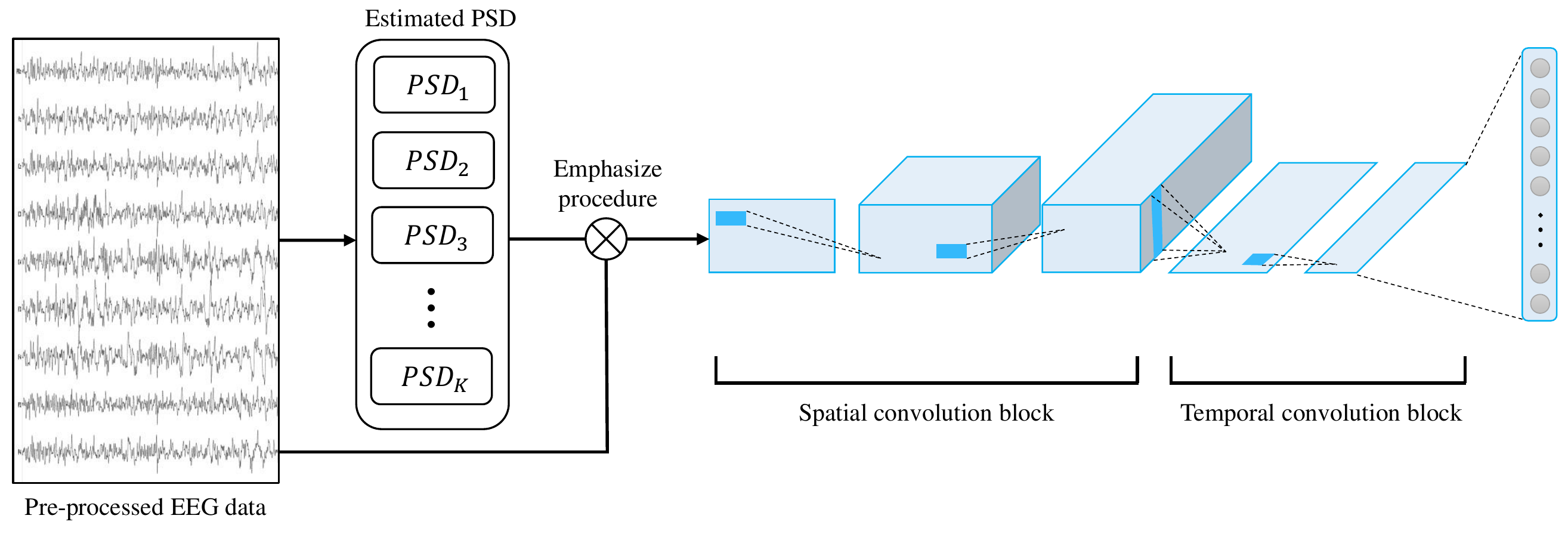}
\caption{The overview of proposed method. The total number of EEG channels is represented by \textit{K}. Emphasized EEG data generates significant features through the spatial convolution block and the temporal convolution block.}
\end{figure*}

In this work, we used the common features from MI. And we proposed a MI-related frequency domain emphasize method to increase MI classification performance. Also, we designed a deep learning architecture that considers both frequency and spatial information. To emphasize the MI-related frequency domain, we calculated the power spectral density (PSD) from the EEG signal \cite{lee2020frontal}. PSD is one of the widely used methods to calculate the power of each frequency range. The proposed method aims to extract the common MI features of users and classify them based on the emphasis on the frequency domain that induces the strongest response in the raw EEG signals. In order to extract the features by considering the frequency and spatial information, we used the convolutional neural network. We evaluated the session-to-session transfer performance using the proposed method.

\section{Methods}

\subsection{Dataset}

We used 2020 International BCI Competition dataset from Korea university (http://brain.korea.ac.kr/bci2020/competition. php). This dataset contains EEG data from 15 healthy subjects who performed MI task. EEG signals were collected using Ag/AgCl 60 electrodes under 10-20 international system by 250 Hz sampling rate. The MI task comprised of 3-class, namely `Cylindrical', `Spherical', and `Lumbrical'. There a total of 300 trials per subjects that are divided into training set and validation set 150 trials, respectively. The scenario of MI task was consist of three-stage, that are relaxation, preparation/cue, and motor imagery, respectively.

\subsection{Data Analysis}
The pre-processing was performed using BBCI toolbox and openBMI\cite{lee2019eeg} with MATLAB 2019a (MathWorks Inc., USA). The dataset was band-pass filtered between [8-30] Hz, which corresponding to ERD/ERS, using Hamming-windowed zero phase finite impulse response (FIR) filters with optimized order (N = 30). We utilized the sliding window method as a data augmentation technique, 2 s length with 50\% overlapping, to increase the number of dataset.

Each trial consists of a total time length of 10 s, of which the MI task was performed corresponds to data for 4 s, from 6 s to 10 s. We calculated the PSD to confirm significant changes in the MI-related frequency domain induced by the brain signal while the user performed the MI tasks. PSD is widely known as a method for extracting the activity of EEG signal by frequency range that want to measure.

\begin{table}[t!]
\small
\caption{Architecture Design of Proposed Method}
\renewcommand{\arraystretch}{1.3}
\resizebox{\columnwidth}{!}{%
\begin{tabular}{ccccc}
\hline
\textbf{Layer} & \textbf{Type} & \textbf{Output shape}    & \textbf{Kernel size} & \textbf{Stride} \\ \hline
1              & Convolution   & 20 × N of channels × 469 & 1 × 32               & 1 × 1           \\
2              & Convolution   & 40 × N of channels × 438 & 1 × 32               & 1 × 1           \\
3              & Convolution   & 40 × 1 × 438             & N of channels × 1    & 1 × 1           \\
4              & Max Pooling   & 40 × 1 × 87              & 1 × 5                & 1 × 5           \\
5              & Dropout (0.5) & -                        & -                    & -               \\
6              & Convolution   & 80 × 1 × 56              & 1 × 32               & 1 × 1           \\
7              & Max Pooling   & 80 × 1 × 11              & 1 × 5                & 1 × 5           \\
8              & Flatten       & 1 × 880                  & -                    & -               \\
9              & Softmax       & 1 × 3                    & -                    & -               \\ \hline
\end{tabular}}
\end{table}

\subsection{Proposed Method}
In this study, PSD was used to calculate the power of the EEG signal in the frequency domain associated with the MI. Subsequently, convolutional neural network (CNN) was used to decode the modified MI data as shown in Fig. 1. We tried to emphasize the spatial area strongly connected to the MI in the brain area. First, we defined an equation for generating modified data by emphasizing MI-related frequency ranges based on PSD:

\begin{align}
PSD=\left \{PSD_{K} \mid 1 \le K \le N of channels \right \}
\end{align}

\begin{equation} PSD_{K} = 10*log_{10}(2\int_{f_1}^{f_2} |\hat{x}(2\pi f)|^2 df) \end{equation}
, where X represent the result of fast fourier transform \cite{bulthoff2003biologically} between the frequency $f_1$, $f_2$. This process is repeated $K$ times, which indicates the total number of channels. The raw EEG data $D$ was modified with the element-wise product process, $X=D \odot PSD_K$ which leads to a focus on the frequency range where we want to emphasize. In order to extract significant features from focused data $X$, we designed the deep learning architecture represented in Table I. The proposed deep learning architecture consisted of two convolution layers and one max pooling layer. A dropout layer is adopted to prevent the overfitting problem due to a small number of the training dataset, and the dropout parameter was selected as 0.5. 

\section{Results and Discussions}
\subsection{Data Analysis}

We investigated the changes of power in the frequency domain through PSD. As shown in Fig. 2, we can infer the significant frequency range during the user performing MI task. It represents significant power values in alpha and beta frequency range, frequencies related to MI. This result means that the MI has been performed properly by the user, and this phenomenon can be a reference to the use of emphasizing the EEG data based on it.

\begin{figure}[t!]
\includegraphics[width=\columnwidth]{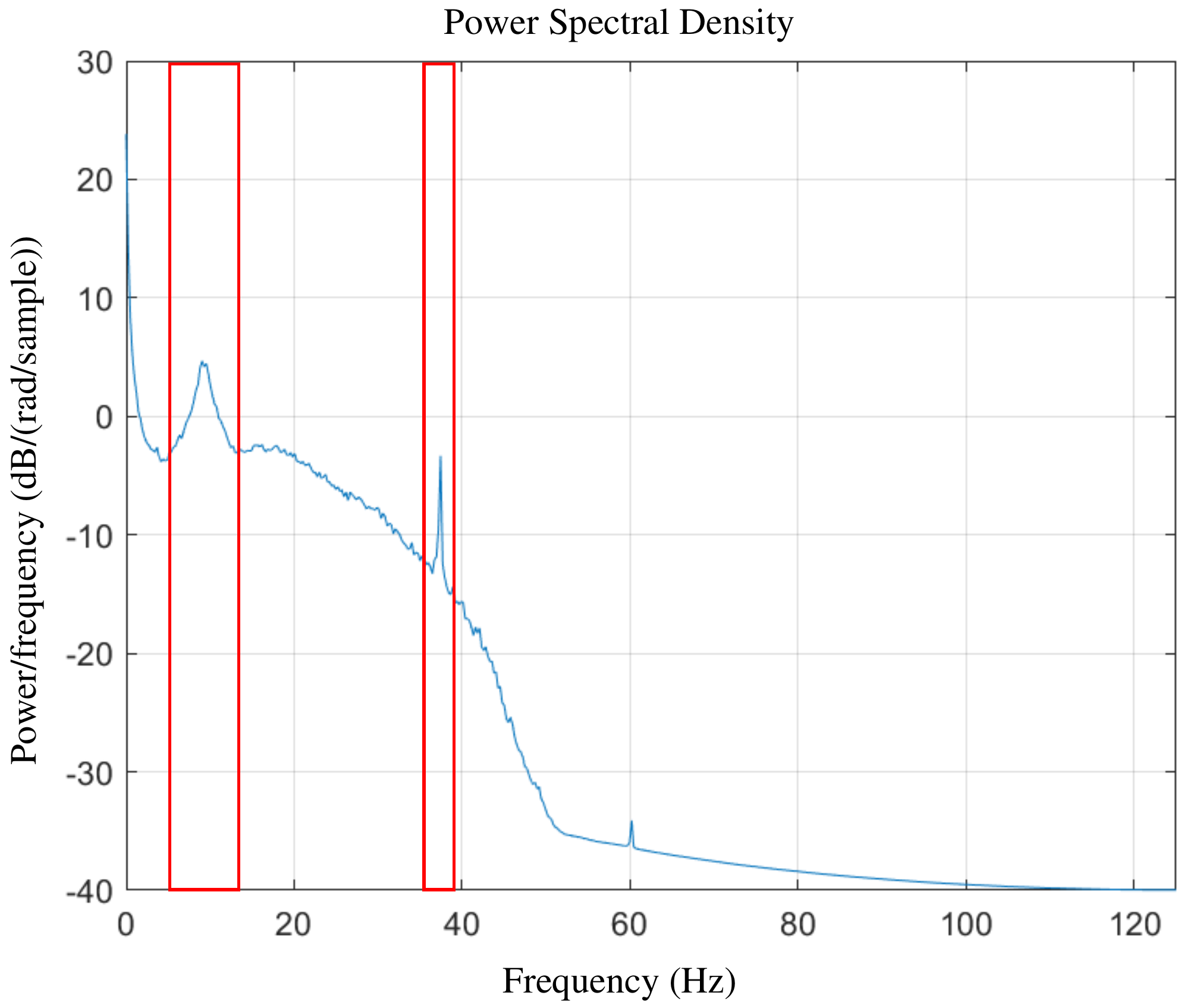}
\caption{The result of power spectral density in motor imagery. This result indicates which frequency range is significant when the user performing the MI task for a representative subject named sub2.}
\end{figure}

\subsection{Performance Evaluation}\label{AA}
We evaluated the competition dataset with proposed method and baseline methods to varify the possibility of controlling the device using non-invasive BCI approach. We used DeepConvNet \cite{schirrmeister2017deep} and EEGNet \cite{lawhern2018eegnet} as baseline to compare the classification performance with proposed method. In this study, we divided dataset into two perspectives: inter-session and intra-session condition to derive classification performance. 

\subsubsection{Intra-Session Condition Classification Performance}
Intra-session condition is the classification of MI performed within one session, and we performed the 5-fold cross validation with the validation set of a given dataset. Table II shows the performance comparison using the baseline methods and proposed method under intra-session condition. The average classification performance using proposed method with MI dataset was 69.67\%, with performance differences ranging from at least 50.12\% per subject to up to 96.67\%. The highest classification performance was recorded when using proposed method, and the average performance was 2.58\% higher than DeepConvNet and 5.15\% higher than EEGNet. The proposed method showed the best performance, and DeepConvNet recorded better performance between EEGNet and DeepConvNet. This result can be interpreted that the data augmentation method using the sliding window collected enough training data and based on this, DeepConvNet consisting of a deeper layer recorded better performance. The proposed method is a shallow network architecture rather than DeepConvNet, but we can guess that it performs better because data is emphasized by channel based on PSD.

\begin{table}[t!]
\caption{Performance Comparison using The Conventional Methods under Intra-Session Environment}
\scriptsize
\renewcommand{\arraystretch}{1.25}
\resizebox{\columnwidth}{!}{
\begin{tabular}{lccc}
\hline
                & \textbf{EEGNet\cite{lawhern2018eegnet}}                                         & \textbf{DeepConvNet\cite{schirrmeister2017deep}}                                    & \textbf{Proposed Method}                                \\ \hline
\textbf{Sub 1}  & \begin{tabular}[c]{@{}c@{}}78.15\% ($\pm$ 3.12)\end{tabular} & \begin{tabular}[c]{@{}c@{}}80.97\% ($\pm$ 3.41)\end{tabular} & \begin{tabular}[c]{@{}c@{}}81.13\% ($\pm$ 2.97)\end{tabular} \\
\textbf{Sub 2}  & \begin{tabular}[c]{@{}c@{}}96.48\% ($\pm$ 1.44)\end{tabular} & \begin{tabular}[c]{@{}c@{}}96.25\% ($\pm$ 1.98)\end{tabular} & \begin{tabular}[c]{@{}c@{}}96.88\% ($\pm$ 1.54)\end{tabular} \\
\textbf{Sub 3}  & \begin{tabular}[c]{@{}c@{}}65.01\% ($\pm$ 4.79)\end{tabular} & \begin{tabular}[c]{@{}c@{}}71.17\% ($\pm$ 1.47)\end{tabular} & \begin{tabular}[c]{@{}c@{}}72.35\% ($\pm$ 3.89)\end{tabular} \\
\textbf{Sub 4}  & \begin{tabular}[c]{@{}c@{}}84.72\% ($\pm$2.21)\end{tabular} & \begin{tabular}[c]{@{}c@{}}86.07\% ($\pm$ 3.37)\end{tabular} & \begin{tabular}[c]{@{}c@{}}90.17\% ($\pm$ 3.14)\end{tabular} \\
\textbf{Sub 5}  & \begin{tabular}[c]{@{}c@{}}58.05\% ($\pm$ 2.91)\end{tabular} & \begin{tabular}[c]{@{}c@{}}53.89\% ($\pm$ 3.13)\end{tabular} & \begin{tabular}[c]{@{}c@{}}67.87\% ($\pm$ 2.78)\end{tabular} \\
\textbf{Sub 6}  & \begin{tabular}[c]{@{}c@{}}71.95\% ($\pm$ 4.32)\end{tabular} & \begin{tabular}[c]{@{}c@{}}69.43\% ($\pm$ 2.74)\end{tabular} & \begin{tabular}[c]{@{}c@{}}83.22\% ($\pm$ 2.18)\end{tabular} \\
\textbf{Sub 7}  & \begin{tabular}[c]{@{}c@{}}47.40\% ($\pm$ 6.11)\end{tabular} & \begin{tabular}[c]{@{}c@{}}57.64\% ($\pm$ 4.55)\end{tabular} & \begin{tabular}[c]{@{}c@{}}58.64\% ($\pm$ 3.97)\end{tabular} \\
\textbf{Sub 8}  & \begin{tabular}[c]{@{}c@{}}53.89\% ($\pm$ 3.89)\end{tabular} & \begin{tabular}[c]{@{}c@{}}48.47\% ($\pm$ 3.37)\end{tabular} & \begin{tabular}[c]{@{}c@{}}64.31\% ($\pm$ 2.77)\end{tabular} \\
\textbf{Sub 9}  & \begin{tabular}[c]{@{}c@{}}43.89\% ($\pm$ 3.97)\end{tabular} & \begin{tabular}[c]{@{}c@{}}47.11\% ($\pm$ 2.36)\end{tabular} & \begin{tabular}[c]{@{}c@{}}50.12\% ($\pm$ 3.85)\end{tabular} \\
\textbf{Sub 10} & \begin{tabular}[c]{@{}c@{}}64.44\% ($\pm$ 1.36)\end{tabular} & \begin{tabular}[c]{@{}c@{}}68.35\% ($\pm$ 1.19)\end{tabular} & \begin{tabular}[c]{@{}c@{}}74.92\% ($\pm$ 2.30)\end{tabular} \\
\textbf{Sub 11} & \begin{tabular}[c]{@{}c@{}}50.83\% ($\pm$ 1.98)\end{tabular} & \begin{tabular}[c]{@{}c@{}}61.22\% ($\pm$ 2.30)\end{tabular} & \begin{tabular}[c]{@{}c@{}}59.55\% ($\pm$ 1.25)\end{tabular} \\
\textbf{Sub 12} & \begin{tabular}[c]{@{}c@{}}73.89\% ($\pm$ 2.85)\end{tabular} & \begin{tabular}[c]{@{}c@{}}73.17\% ($\pm$ 3.55)\end{tabular} & \begin{tabular}[c]{@{}c@{}}75.78\% ($\pm$ 1.51)\end{tabular} \\
\textbf{Sub 13} & \begin{tabular}[c]{@{}c@{}}47.78\% ($\pm$ 3.85)\end{tabular} & \begin{tabular}[c]{@{}c@{}}55.62\% ($\pm$ 4.22)\end{tabular} & \begin{tabular}[c]{@{}c@{}}57.21\% ($\pm$ 3.86)\end{tabular} \\
\textbf{Sub 14} & \begin{tabular}[c]{@{}c@{}}66.39\% ($\pm$ 3.63)\end{tabular} & \begin{tabular}[c]{@{}c@{}}69.18\% ($\pm$ 5.44)\end{tabular} & \begin{tabular}[c]{@{}c@{}}73.85\% ($\pm$ 4.81)\end{tabular} \\
\textbf{Sub 15} & \begin{tabular}[c]{@{}c@{}}65.42\% ($\pm$ 1.99)\end{tabular} & \begin{tabular}[c]{@{}c@{}}67.45\% ($\pm$ 2.14)\end{tabular} & \begin{tabular}[c]{@{}c@{}}66.14\% ($\pm$ 2.07)\end{tabular} \\
\textbf{Avg.}   & \begin{tabular}[c]{@{}c@{}}64.53\% ($\pm$ 13.95)\end{tabular} & \begin{tabular}[c]{@{}c@{}}67.07\% ($\pm$ 12.85)\end{tabular} & \begin{tabular}[c]{@{}c@{}}69.68\% ($\pm$ 10.10)\end{tabular} \\ \hline
\end{tabular}
}
\end{table}

\begin{table}[t!]
\caption{Performance Comparison using The Conventional Methods under Inter-Session Environment}
\scriptsize
\renewcommand{\arraystretch}{1.25}
\resizebox{\columnwidth}{!}{
\begin{tabular}{lccc}
\hline
                & \textbf{EEGNet\cite{lawhern2018eegnet}}                                         & \textbf{DeepConvNet\cite{schirrmeister2017deep}}                                    & \textbf{Proposed Method}                                \\ \hline
\textbf{Sub 1}  & \begin{tabular}[c]{@{}c@{}}31.33\% ($\pm$ 4.12)\end{tabular} & \begin{tabular}[c]{@{}c@{}}35.58\% ($\pm$ 3.48)\end{tabular} & \begin{tabular}[c]{@{}c@{}}41.55\% ($\pm$ 4.01)\end{tabular} \\
\textbf{Sub 2}  & \begin{tabular}[c]{@{}c@{}}33.47\% ($\pm$ 2.35)\end{tabular} & \begin{tabular}[c]{@{}c@{}}47.14\% ($\pm$ 3.88)\end{tabular} & \begin{tabular}[c]{@{}c@{}}50.63\% ($\pm$ 3.21)\end{tabular} \\
\textbf{Sub 3}  & \begin{tabular}[c]{@{}c@{}}40.84\% ($\pm$ 5.12)\end{tabular} & \begin{tabular}[c]{@{}c@{}}38.34\% ($\pm$ 4.57)\end{tabular} & \begin{tabular}[c]{@{}c@{}}42.27\% ($\pm$ 4.11)\end{tabular} \\
\textbf{Sub 4}  & \begin{tabular}[c]{@{}c@{}}76.24\% ($\pm$ 2.48)\end{tabular} & \begin{tabular}[c]{@{}c@{}}78.18\% ($\pm$ 2.01)\end{tabular} & \begin{tabular}[c]{@{}c@{}}82.69\% ($\pm$ 3.15)\end{tabular} \\
\textbf{Sub 5}  & \begin{tabular}[c]{@{}c@{}}41.01\% ($\pm$ 4.25)\end{tabular} & \begin{tabular}[c]{@{}c@{}}51.09\% ($\pm$ 3.12)\end{tabular} & \begin{tabular}[c]{@{}c@{}}54.41\% ($\pm$ 2.85)\end{tabular} \\
\textbf{Sub 6}  & \begin{tabular}[c]{@{}c@{}}55.39\% ($\pm$ 2.14)\end{tabular} & \begin{tabular}[c]{@{}c@{}}64.77\% ($\pm$ 3.85)\end{tabular} & \begin{tabular}[c]{@{}c@{}}68.47\% ($\pm$ 1.90)\end{tabular} \\
\textbf{Sub 7}  & \begin{tabular}[c]{@{}c@{}}48.07\% ($\pm$ 1.24)\end{tabular} & \begin{tabular}[c]{@{}c@{}}52.89\% ($\pm$ 3.44)\end{tabular} & \begin{tabular}[c]{@{}c@{}}52.27\% ($\pm$ 3.15)\end{tabular} \\
\textbf{Sub 8}  & \begin{tabular}[c]{@{}c@{}}33.54\% ($\pm$ 4.18)\end{tabular} & \begin{tabular}[c]{@{}c@{}}37.81\% ($\pm$ 4.07)\end{tabular} & \begin{tabular}[c]{@{}c@{}}36.88\% ($\pm$ 4.55)\end{tabular} \\
\textbf{Sub 9}  & \begin{tabular}[c]{@{}c@{}}39.71\% ($\pm$ 3.64)\end{tabular} & \begin{tabular}[c]{@{}c@{}}38.14\% ($\pm$ 2.58)\end{tabular} & \begin{tabular}[c]{@{}c@{}}42.08\% ($\pm$ 3.27)\end{tabular} \\
\textbf{Sub 10} & \begin{tabular}[c]{@{}c@{}}37.24\% ($\pm$ 1.63)\end{tabular} & \begin{tabular}[c]{@{}c@{}}37.51\% ($\pm$ 2.88)\end{tabular} & \begin{tabular}[c]{@{}c@{}}38.91\% ($\pm$ 1.58)\end{tabular} \\
\textbf{Sub 11} & \begin{tabular}[c]{@{}c@{}}49.89\% ($\pm$ 2.74)\end{tabular} & \begin{tabular}[c]{@{}c@{}}54.89\% ($\pm$ 3.41)\end{tabular} & \begin{tabular}[c]{@{}c@{}}49.88\% ($\pm$ 3.22)\end{tabular} \\
\textbf{Sub 12} & \begin{tabular}[c]{@{}c@{}}55.80\% ($\pm$ 3.47)\end{tabular} & \begin{tabular}[c]{@{}c@{}}62.04\% ($\pm$ 2.15)\end{tabular} & \begin{tabular}[c]{@{}c@{}}65.17\% ($\pm$ 2.01)\end{tabular} \\
\textbf{Sub 13} & \begin{tabular}[c]{@{}c@{}}42.48\% ($\pm$ 2.33)\end{tabular} & \begin{tabular}[c]{@{}c@{}}49.88\% ($\pm$ 2.54)\end{tabular} & \begin{tabular}[c]{@{}c@{}}48.46\% ($\pm$ 3.01)\end{tabular} \\
\textbf{Sub 14} & \begin{tabular}[c]{@{}c@{}}50.91\% ($\pm$ 1.98)\end{tabular} & \begin{tabular}[c]{@{}c@{}}51.24\% ($\pm$ 2.55)\end{tabular} & \begin{tabular}[c]{@{}c@{}}54.10\% ($\pm$ 1.81)\end{tabular} \\
\textbf{Sub 15} & \begin{tabular}[c]{@{}c@{}}56.17\% ($\pm$ 2.39)\end{tabular} & \begin{tabular}[c]{@{}c@{}}62.23\% ($\pm$ 1.99)\end{tabular} & \begin{tabular}[c]{@{}c@{}}63.58\% ($\pm$ 2.07)\end{tabular} \\
\textbf{Avg.}   & \begin{tabular}[c]{@{}c@{}}46.14\% ($\pm$ 11.41)\end{tabular} & \begin{tabular}[c]{@{}c@{}}50.78\% ($\pm$ 11.91)\end{tabular} & \begin{tabular}[c]{@{}c@{}}52.76\% ($\pm$ 12.23)\end{tabular} \\ \hline
\end{tabular}
}
\end{table}

\subsubsection{Inter-Session Condition Classification Performance}
The inter-session condition means dividing data generated from two sessions from one user into training data and validation data, respectively. As shown in Table III, the proposed method derived the highest classification performance compared with the baseline methods. And the average classification performance when using the proposed method was 1.98\% higher than DeepConvNet and 6.62\% higher than EEGNet. These results allow us to infer that the deep learning architecture extracts significant features when using methods that emphasize the MI-related frequency domain. In the proposed method, the performance in inter-session condition is lower than the performance in intra-session condition, but has a similar tendency. Baseline methods extract features only through deep learning architecture without considering the significant features of MI, so we can infer that they derive lower performance than proposed methods.

\section{Conclusions and Future Works}
In this study, we designed a deep learning network and applied MI corresponded frequency domain emphasize method to decode the MI dataset related to grasp including user intention. First, we used PSD to confirm the significant frequency range in the MI dataset. Subsequently, the frequency range with a large PSD was considered a significant frequency range in the MI, and this was used to modify the dataset. The modified dataset was emphasized by channel using a significant frequency range, through which the deep learning architecture we proposed was constructed. The deep learning architecture was constructed using the convolution layer and the significant features were extracted with spatial information. As a result, we derived 3.10\% and 8.46\% higher performance than the baseline method EEGNet and DeepConvNet, respectively. 

\section*{Acknowledgment}

The authors thank to Mr. G.-H. Shin and Ms. D.-Y Lee for useful discussion.\\

\bibliographystyle{IEEEtran}
\bibliography{ref}

\begin{thebibliography}{10}
\providecommand{\url}[1]{#1}
\csname url@samestyle\endcsname
\providecommand{\newblock}{\relax}
\providecommand{\bibinfo}[2]{#2}
\providecommand{\BIBentrySTDinterwordspacing}{\spaceskip=0pt\relax}
\providecommand{\BIBentryALTinterwordstretchfactor}{4}
\providecommand{\BIBentryALTinterwordspacing}{\spaceskip=\fontdimen2\font plus
\BIBentryALTinterwordstretchfactor\fontdimen3\font minus
  \fontdimen4\font\relax}
\providecommand{\BIBforeignlanguage}[2]{{%
\expandafter\ifx\csname l@#1\endcsname\relax
\typeout{** WARNING: IEEEtran.bst: No hyphenation pattern has been}%
\typeout{** loaded for the language `#1'. Using the pattern for}%
\typeout{** the default language instead.}%
\else
\language=\csname l@#1\endcsname
\fi
#2}}
\providecommand{\BIBdecl}{\relax}
\BIBdecl

\bibitem{vaughan2003brain}
{T. M. Vaughan, W. J. Heetderks, L. J. Trejo, W. Z. Rymer, M. Weinrich, M. M.
  Moore, A. Kübler, B. H. Dobkin, N. Birbaumer, E. Donchin, E. W. Wolpaw, J.
  R. Wolpaw}, ``{Brain-computer interface technology: a review of the Second
  International Meeting},'' \emph{IEEE Trans. Neural Syst. Rehabil. Eng.},
  vol.~11, pp. 94--109, Jun. 2003.

\bibitem{chen2016high}
{Y. Chen, A. D. Atnafu, I. Schlattner, W. T. Weldtsadik, M.-C. Roh, H.-J. Kim,
  S.-W. Lee, B. Blankertz, and S. Fazli}, ``{A high-security {EEG}-based login
  system with RSVP stimuli and dry electrodes},'' \emph{IEEE Trans. Inf.
  Forensics Secur.}, vol.~11, pp. 2635--2647, Jun. 2016.

\bibitem{lee2017network}
M.~Lee, R.~D. Sanders, S.-K. Yeom, D.-O. Won, K.-S. Seo, H.~J. Kim, G.~Tononi,
  and S.-W. Lee, ``Network properties in transitions of consciousness during
  propofol-induced sedation,'' \emph{Sci. Rep.}, vol.~7, no.~1, pp. 1--13, Dec.
  2017.

\bibitem{jeong2019trajectory}
{J.-H. Jeong, K.-H. Shim, J.-H. Cho, and S.-W. Lee}, ``{Trajectory decoding of
  arm reaching movement imageries for brain–controlled robot arm system},''
  in \emph{Conf. Proc. IEEE Eng. Med. Biol. Soc. (EMBC)}, Berlin, Germany, Jul.
  2019, pp. 23--27.

\bibitem{he2018brain}
Y.~He, D.~Eguren, J.~M. Azor{\'\i}n, R.~G. Grossman, T.~P. Luu, and J.~L.
  Contreras-Vidal, ``Brain--machine interfaces for controlling lower-limb
  powered robotic systems,'' \emph{J. Neural. Eng.}, vol.~15, no.~2, p. 021004,
  Feb. 2018.

\bibitem{lee2020continuous}
D.-H. Lee, J.-H. Jeong, K.~Kim, B.-W. Yu, and S.-W. Lee, ``Continuous {EEG}
  decoding of pilots’ mental states using multiple feature block-based
  convolutional neural network,'' \emph{IEEE Access}, vol.~8, pp.
  121\,929--121\,941, Jul. 2020.

\bibitem{park2016movement}
K.-H. Park and S.-W. Lee, ``Movement intention decoding based on deep learning
  for multiuser myoelectric interfaces,'' in \emph{Int. Winter Conf.
  Brain-Computer Interface (BCI)}, Jeongseon, Republic of Korea, Feb. 2016, pp.
  1--2.

\bibitem{suk2014predicting}
H.-I. Suk, S.~Fazli, J.~Mehnert, K.-R. M{\"u}ller, and S.-W. Lee, ``Predicting
  {BCI} subject performance using probabilistic spatio-temporal filters,''
  \emph{PloS one}, vol.~9, no.~2, p. e87056, Feb. 2014.

\bibitem{kwak2017convolutional}
{N.-S. Kwak, K. R. M{\"u}ller, and S.-W. Lee}, ``{A convolutional neural
  network for steady state visual evoked potential classification under
  ambulatory environment},'' \emph{PLoS One}, vol.~12, p. e0172578, Feb. 2017.

\bibitem{chen2018control}
X.~Chen, B.~Zhao, Y.~Wang, S.~Xu, and X.~Gao, ``Control of a 7-{DOF} robotic
  arm system with an {SSVEP}-based {BCI},'' \emph{Int. J. Neural Syst.},
  vol.~28, no.~8, p. 1850018, Oct. 2018.

\bibitem{won2015effect}
{D.-O. Won, H.-J. Hwang, S. D{\"a}hne, K.-R. M{\"u}ller, and S.-W. Lee},
  ``{Effect of higher frequency on the classification of steady-state visual
  evoked potentials},'' \emph{J. Neural Eng.}, vol.~13, p. 016014, Dec. 2015.

\bibitem{yeom2014efficient}
{S.-K. Yeom, S. Fazli, K. R. M{\"u}ller, and S.-W. Lee}, ``{An efficient
  ERP-based brain-computer interface using random set presentation and face
  familiarity},'' \emph{PLoS One}, vol.~9, p. e111157, Nov. 2014.

\bibitem{lee2018high}
M.-H. Lee, J.~Williamson, D.-O. Won, S.~Fazli, and S.-W. Lee, ``A high
  performance spelling system based on {EEG}-{EOG} signals with visual
  feedback,'' \emph{IEEE Trans. Neural Syst. Rehabil. Eng.}, vol.~26, no.~7,
  pp. 1443--1459, Jul. 2018.

\bibitem{kim2014decoding}
{J.-H. Kim, F. Bie{\ss}mann, and S.-W. Lee}, ``{Decoding three-dimensional
  trajectory of executed and imagined arm movements from electroencephalogram
  signals},'' \emph{IEEE Trans. Neural Syst. Rehabil. Eng.}, vol.~23, pp.
  867--876, Dec. 2014.

\bibitem{ofner2014using}
P.~Ofner and G.~R. M{\"u}ller-Putz, ``Using a noninvasive decoding method to
  classify rhythmic movement imaginations of the arm in two planes,''
  \emph{IEEE transactions on biomedical engineering}, vol.~62, no.~3, pp.
  972--981, Mar. 2014.

\bibitem{kam2013non}
{T.-E. Kam, H.-I. Suk, and S.-W. Lee}, ``{Non-homogeneous spatial filter
  optimization for ElectroEncephaloGram (EEG)-based motor imagery
  classification},'' \emph{Neurocomputing}, vol. 108, pp. 58--68, May 2013.

\bibitem{won2017motion}
{D.-O. Won, H.-J. Hwang, D.-M. Kim, K.-R. M{\"u}ller, and S.-W. Lee},
  ``{Motion-based rapid serial visual presentation for gaze-independent
  brain-computer interfaces},'' \emph{IEEE Trans. Neural Syst. Rehabil. Eng.},
  vol.~26, pp. 334--343, Aug. 2017.

\bibitem{meng2016noninvasive}
{J. Meng, S. Zhang, A. Bekyo, J. Olsoe, B. Baxter, and B. He}, ``{Noninvasive
  electroencephalogram based control of a robotic arm for reach and grasp
  tasks},'' \emph{Sci. Rep.}, vol.~6, pp. 1--15, Dec. 2016.

\bibitem{shiman2017classification}
F.~Shiman, E.~L{\'o}pez-Larraz, A.~Sarasola-Sanz, N.~Irastorza-Landa,
  M.~Sp{\"u}ler, N.~Birbaumer, and A.~Ramos-Murguialday, ``Classification of
  different reaching movements from the same limb using {EEG},'' \emph{Journal
  of neural engineering}, vol.~14, no.~4, p. 046018, Jun. 2017.

\bibitem{spataro2017reaching}
R.~Spataro, A.~Chella, B.~Allison, M.~Giardina, R.~Sorbello, S.~Tramonte,
  C.~Guger, and V.~La~Bella, ``Reaching and grasping a glass of water by
  locked-in {ALS} patients through a {BCI}-controlled humanoid robot,''
  \emph{Front. Hum. Neurosci.}, vol.~11, p.~68, Mar. 2017.

\bibitem{kim2016commanding}
{K.-T. Kim, H.-I. Suk, and S.-W. Lee}, ``{Commanding a brain-controlled
  wheelchair using steady-state somatosensory evoked potentials},'' \emph{IEEE
  Trans. Neural Syst. Rehabil. Eng.}, vol.~26, pp. 654--665, Aug. 2018.

\bibitem{jeong2020brain}
J.-H. Jeong, K.-H. Shim, D.-J. Kim, and S.-W. Lee, ``Brain-controlled robotic
  arm system based on multi-directional {CNN}-{B}i{LSTM} network using {EEG}
  signals,'' \emph{IEEE Trans. Neural Syst. Rehabil. Eng.}, vol.~28, no.~5, pp.
  1226--1238, Mar. 2020.

\bibitem{korik2018decoding}
{A. Korik, R. Sosnik, N. Siddique, and D. Coyle}, ``{Decoding imagined 3D hand
  movement trajectories from EEG: evidence to support the use of mu, beta, and
  low gamma oscillations},'' \emph{Front. Neurosci.}, vol.~12, p. 130, Mar.
  2018.

\bibitem{lee2015subject}
{M.-H. Lee, S. Fazli, J. Mehnert, and S.-W. Lee}, ``{Subject-dependent
  classification for robust idle state detection using multi-modal neuroimaging
  and data-fusion techniques in BCI},'' \emph{Pattern Recognit.}, vol.~48, pp.
  2725--2737, Aug. 2015.

\bibitem{zhang2019survey}
X.~Zhang, L.~Yao, X.~Wang, J.~Monaghan, D.~Mcalpine, and Y.~Zhang, ``A survey
  on deep learning based brain computer interface: Recent advances and new
  frontiers,'' \emph{arXiv preprint arXiv:1905.04149}, Oct. 2019.

\bibitem{raza2019covariate}
H.~Raza, D.~Rathee, S.-M. Zhou, H.~Cecotti, and G.~Prasad, ``Covariate shift
  estimation based adaptive ensemble learning for handling non-stationarity in
  motor imagery related {EEG}-based brain-computer interface,''
  \emph{Neurocomputing}, vol. 343, pp. 154--166, May 2019.

\bibitem{lee2020frontal}
M.~Lee, G.-H. Shin, and S.-W. Lee, ``Frontal {EEG} asymmetry of emotion for the
  same auditory stimulus,'' \emph{IEEE Access}, vol.~8, pp. 107\,200--107\,213,
  Jun. 2020.

\bibitem{lee2019eeg}
M.-H. Lee, O.-Y. Kwon, Y.-J. Kim, H.-K. Kim, Y.-E. Lee, J.~Williamson,
  S.~Fazli, and S.-W. Lee, ``{EEG} dataset and {O}pen{BMI} toolbox for three
  {BCI} paradigms: an investigation into {BCI} illiteracy,''
  \emph{GigaScience}, vol.~8, no.~5, p. giz002, Jan. 2019.

\bibitem{bulthoff2003biologically}
H.~H. B{\"u}lthoff, S.-W. Lee, T.~Poggio, and C.~Wallraven, \emph{Biologically
  motivated computer vision}.\hskip 1em plus 0.5em minus 0.4em\relax NY:
  Springer-Verlag, 2003.

\bibitem{schirrmeister2017deep}
{R. T. Schirrmeister, J. T. Springenberg, L. D. J. Fiederer, M. Glasstetter, K.
  Eggensperger, M. Tangermann, F. Hutter, W. Burgard, and T. Ball}, ``Deep
  learning with convolutional neural networks for {EEG} decoding and
  visualization,'' \emph{Hum. Brain Mapp.}, vol.~38, no.~11, pp. 5391--5420,
  Aug. 2017.

\bibitem{lawhern2018eegnet}
V.~J. Lawhern, A.~J. Solon, N.~R. Waytowich, S.~M. Gordon, C.~P. Hung, and
  B.~J. Lance, ``{EEGN}et: a compact convolutional neural network for
  {EEG}-based brain--computer interfaces,'' \emph{J. Neural Eng.}, vol.~15,
  no.~5, p. 056013, Jul. 2018.

\end{thebibliography}

\end{document}